\documentclass[11pt,twoside]{article}
\usepackage{asp2004}
\usepackage{psfig}
\usepackage{epsf}
\usepackage{graphics}
\usepackage{graphicx}
\usepackage{lscape}
\def 	  \be 		{\begin{equation}}
\def	  \ee		{\end{equation}}

\def\edcomment#1{\iffalse\marginpar{\raggedright\sl#1\/}\else\relax\fi}
\reversemarginpar

\begin{document}
\title{Grain Alignment in Molecular Clouds}
\author{A. Lazarian and J. Cho}
\affil{University of Wisconsin-Madison: \texttt{lazarian, cho@astro.wisc.edu}}

\begin{abstract}
Polarimetry is one of the most informative techniques of studying magnetic 
fields in molecular clouds. How reliable
the interpretation of the polarization maps in terms of magnetic
fields is the issue that the grain alignment theory addresses. 
 We show that grain
alignment involves several processes acting simultaneously, but
on different time-scales. We explain that rotating dust grains
get substantial magnetic moment that allows them precess fast about
magnetic field lines. As the result, grains preserve their orientation
to magnetic field when the magnetic field direction
fluctuates. We point out to the importance of internal alignment, i.e. the
process forces grain axes to be aligned 
in respect to the grain angular momentum. We show
that subtle quantum effects, in particular relaxation related to nuclear
magnetic moments of atoms composing the grain, brings to live complex 
grain motions, e.g. flips. These flips substantially alter the dynamics of 
grain and limit the applicability of earlier theories that did not account
for them. We also briefly review basic physical processes involved in
the alignment of grain angular momentum in respect to interstellar magnetic 
field. We claim that the bulk of existing observational data is consistent
with the radiative torque alignment mechanism. In particular, we show
that large grains that are known to exist in the cores of molecular clouds may
be aligned by the attenuated external interstellar radiation field.

\end{abstract}
\thispagestyle{plain}
\section{Why do we care?}
\label{sec:1}
The fact that interstellar grains get aligned has been puzzling researchers for more than half a 
century. Very soon after the
discovery of grain alignment by Hall (1949) and Hiltner (1949) it became clear 
that the alignment 
happens in respect to magnetic field. Since that time grain alignment 
stopped to be the issue
of pure scientific curiosity, but became an important link 
connecting polarimetry
observations with the all-important interstellar magnetic fields\footnote{
Additional interest to grain alignment arises from recent attempts to
separate the polarized CMB radiation from the polarized foregrounds (see
Lazarian \& Finkbeiner 2003 for a review).}.

The history of grain alignment ideas is exciting (see review by Lazarian 2003) 
but we do not
have space here to dwell upon it.   
Last 
decade has been marked by a substantial progress in understanding new
 physics associated with grain alignment. The theory has become predictive,
which enables researchers to interpret observational data with more confidence.

 Within this short review we discuss 
the modern understanding
of grain alignment processes applicable to molecular clouds. We discuss 
both internal alignment, i.e. the alignment of grain axes in respect
to grain angular momentum, and the alignment of grain angular momentum
in respect to magnetic field. Due to fast grain precession about magnetic 
field the latter acts as the alignment axis for various alignment mechanisms.
We show that at present the radiative torque alignment is the most
promissing mechanism for explaining the bulk of relevant polarimetry data.
However, we show that other mechanisms also have their nishes.

Recent reviews of the grain alignment theory include Roberge (2004), 
Lazarian (2003)\footnote{The presentation in 
Lazarian (2003) goes beyond molecular cloud
environment and deals with the possibility of alignment in circumstellar 
regions, interplanetary medium,
coma of comets etc. For these regions 
aligned grain have great and yet untapped potential for
studying magnetic fields. The aforementioned review also  deals
 with circular polarization
arising from aligned grains.}. 
Progress in testing theory is covered in Hildebrand (2000), while 
unusual and exciting aspects of grain dynamics
are discussed in Lazarian \& Yan (2004). 
The interested reader may use the reviews above to guide 
her in the vast and exciting
original literature on grain alignment theory. Polarization from aligned
atoms is discussed in a companion paper by Yan \& Lazarian (this volume). 

\section{How does alignment cause polarization?}

Aligned grains absorb more light along their longer direction.
The situation is reversed if grain emission is considered: more emission
emanates in the direction of the longer grain axis. 

Consider polarization arising due to selective extinction of grains first.
For an ensemble of aligned grains the extinction perpendicular 
and parallel to
the direction
of alignment and parallel are different\footnote{According
to Hildebrand \& Dragovan (1995) the best fit of the grain properties
corresponds to oblate grains with the ratio of axis about 2/3.}. Therefore
 that is initially unpolarized starlight acquires polarization while
passing through a volume with aligned grains.
If the extinction in the direction of alignment is $\tau_{\|}$ and in
the perpendicular direction is  $\tau_{\bot}$
one can write the polarization, $P_{abs}$, by selective extinction
 of grains
as 
\begin{equation}
P_{abs}=\frac{e^{-\tau_{\|}}-e^{-\tau_{\bot}}}{e^{-\tau_{\|}}+e^{-\tau_{\bot}}}
\approx -{(\tau_{\|}-\tau_{\bot})}/2~,
\label{Pabs}
\end{equation}
where the latter approximation is valid for $\tau_{\|}-\tau_{\bot}\ll 1$.
To relate the difference of extinction to the properties of aligned grains
one can take into 
account the fact that the extinction is proportional to the product
of the grain density and  their cross sections. If a cloud is composed of 
identical aligned grains
$\tau_{\|}$ and $\tau_{\bot}$ are proportional to the number of grains
along the light path times the corresponding cross sections, which
are, respectively, 
$C_{\|}$ and $C_{\bot}$.

In reality one has to consider additional complications like
incomplete grain alignment, and variations in the direction
of the alignment axis along the line of sight.  
 To obtain 
an adequate description one can (see Roberge \& Lazarian 1999) consider
 an electromagnetic wave propagating along the line of sight
{\mbox{$\hat{\bf z}^{\bf\rm o}$}} axis.
The transfer equations for the Stokes parameters
depend on the cross sections,  $C_{xo}$ and $C_{yo}$, for linearly polarized
waves with the electric vector,  {\mbox{\boldmath$E$}}, 
along the {\mbox{$\hat{\bf x}^{\bf\rm o}$}} and 
{\mbox{$\hat{\bf y}^{\bf\rm o}$}} directions
that are in the plane perpendicular to {\mbox{$\hat{\bf z}^{\bf\rm o}$}}
(see Lee \& Draine 1985).

To calculate  $C_{xo}$ and $C_{yo}$,
one transforms the components of {\mbox{\boldmath$E$}} to
a frame aligned with the principal axes of the grain and
takes the appropriately-weighted sum of the
cross sections, $C_{\|}$ and , $C_{\bot}$ for {\mbox{\boldmath$E$}}
 polarized along the grain
axes.
When the transformation is carried out and the resulting
expressions are averaged over precession angles, one finds (see
transformations in Lee \& Draine 1985 for spheroidal grains and in
Efroimsky 2002 for a general case)
that
the mean cross sections are
\begin{equation}
C_{xo} = C_{avg} + \frac{1}{3}\,R\,\left(C_{\bot}-C_{\|}\right)\,
       \left(1-3\cos^2\zeta\right)~~~,
\label{eq-2_5}
\end{equation}
\begin{equation}
C_{yo} = C_{avg} + \frac{1}{3}\,R\,\left(C_{\bot}-C_{\|}\right)~~~,
\label{eq-2_6}
\end{equation}
where $\zeta$ is the angle between the polarization axis and the 
{\mbox{$\hat{\bf x}^{\bf\rm o}$}} {\mbox{$\hat{\bf y}^{\bf\rm o}$}}
plane;
$C_{avg}\equiv\left(2 C_{\bot}+ C_{\|}\right)/3$ is the effective
cross section for randomly-oriented grains.
To characterize the alignment we used in eq.~(\ref{eq-2_6})
 the Raylegh reduction factor
(Greenberg 1968) $$
R\equiv \langle G(\cos^2\theta) G(\cos^2\beta)\rangle
$$, where angular brackets denote ensemble averaging, $G(x) \equiv 3/2 (x-1/3)$,
 $\theta$ is the angle between the axis of the largest moment of inertia
(henceforth the axis of maximal inertia, see Fig~1) and the magnetic field ${\bf B}$, while
$\theta$ is the angle between the angular momentum ${\bf J}$ and ${\bf B}$. 
 To characterize
${\bf J}$ alignment in grain axes and in respect to magnetic field,
 the
measures ${Q_X\equiv \langle G(\theta)\rangle}$ and 
$Q_J\equiv \langle G(\beta)\rangle$
are used.
Unfortunately, these statistics 
are not independent and therefore $R$ is not equal to $Q_J Q_X$ (see  Roberge
\& Lazarian 1999). This considerably complicates 
the treatment of grain alignment.

Polarization arising from emitting grains can be calculated as follows:
\begin{equation}
P_{em}=\frac{(1-e^{-\tau_{\|}})-(1-e^{-\tau_{\bot}})}{(1-e^{-\tau_{\|}})+
(1-e^{-\tau_{\bot}})}\approx \frac{\tau_{\|}-\tau_{\bot}}
{\tau_{\|}+\tau_{\bot}},
\end{equation}
 where both the optical depths $\tau{\|}$ are $\tau_{\bot}$ were
assumed to be small. Taking into account that 
both $P_{em}$ and $P_{abs}$ are functions
of wavelength $\lambda$ and 
combining eqs.(\ref{Pabs}) and (\ref{Pem}), one gets for
$\tau=(\tau_{\|}+\tau_{\bot})/2$ 
\begin{equation}
P_{em}(\lambda) \approx -P_{abs}(\lambda)/\tau(\lambda)~,
\label{Pem}
\end{equation}
which establishes the relation between polarization in emission and
absorption. The minus sign
in eq~(\ref{Pem})
reflects the fact that emission and absorption polarization are
orthogonal. 
As $P_{abs}$ depends on $R$, $P_{em}$ also depends on $R$.

\section{How complex is grain motion?}

Dynamics of grains in molecular clouds is pretty involved
(see Fig.~1). Grain rotation can
arise from chaotic gaseous bombardment of grain surface and be Brownian, 
or it can arise from 
{\it systematic torques} discovered by Purcell (1975, 1979). 
The most efficient 
among those are torques
arising from H$_2$ formation over grain surface. One can visualize those 
torques imagining a grain
with tiny rocket nozzles ejecting nascent high velocity hydrogen molecules
(see Fig.~1). Indeed, H$_2$ formation is   believed to 
take place over particular
catalytic sites on grain surface. These catalytic sites ejecting molecules
are frequently called 
"Purcell rockets". Even when the surroundings
of  dust grains is mostly molecular, accoring to Purcell (1979) grains can rotate suprathermally, i.e. 
with kinetic energies 
much larger that $kT_{gas}$, due to the variation of the accommodation 
coefficient. Indeed, if 
the temperatures of gas and dust are different, the variations of the
sticking probabilities allow 
parts of the grain to bounce
back impinging gaseous atoms with different efficiencies. In addition, Purcell (1979) identified electron ejection as yet another
process that can drive grain to large angular velocities. All these three
processes are so natural that until very recently it was generally accepted
that all interstellar grains in diffuse interstellar gas {\it must} rotate suprathermally.

\begin{figure}
 \centering \leavevmode
\epsfxsize=1.6in\epsfbox{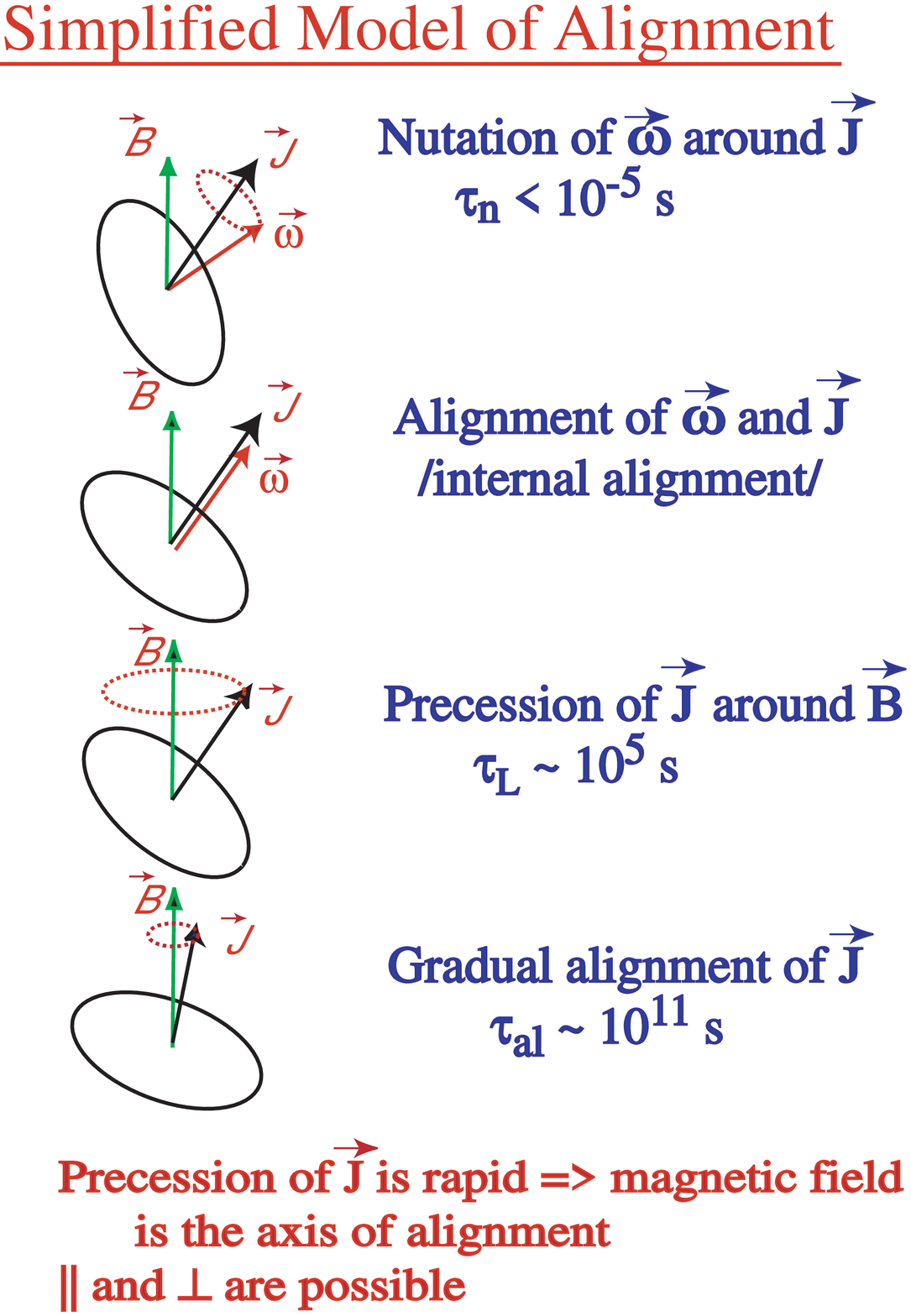}
\hfil
 \epsfxsize=1.8in\epsfbox{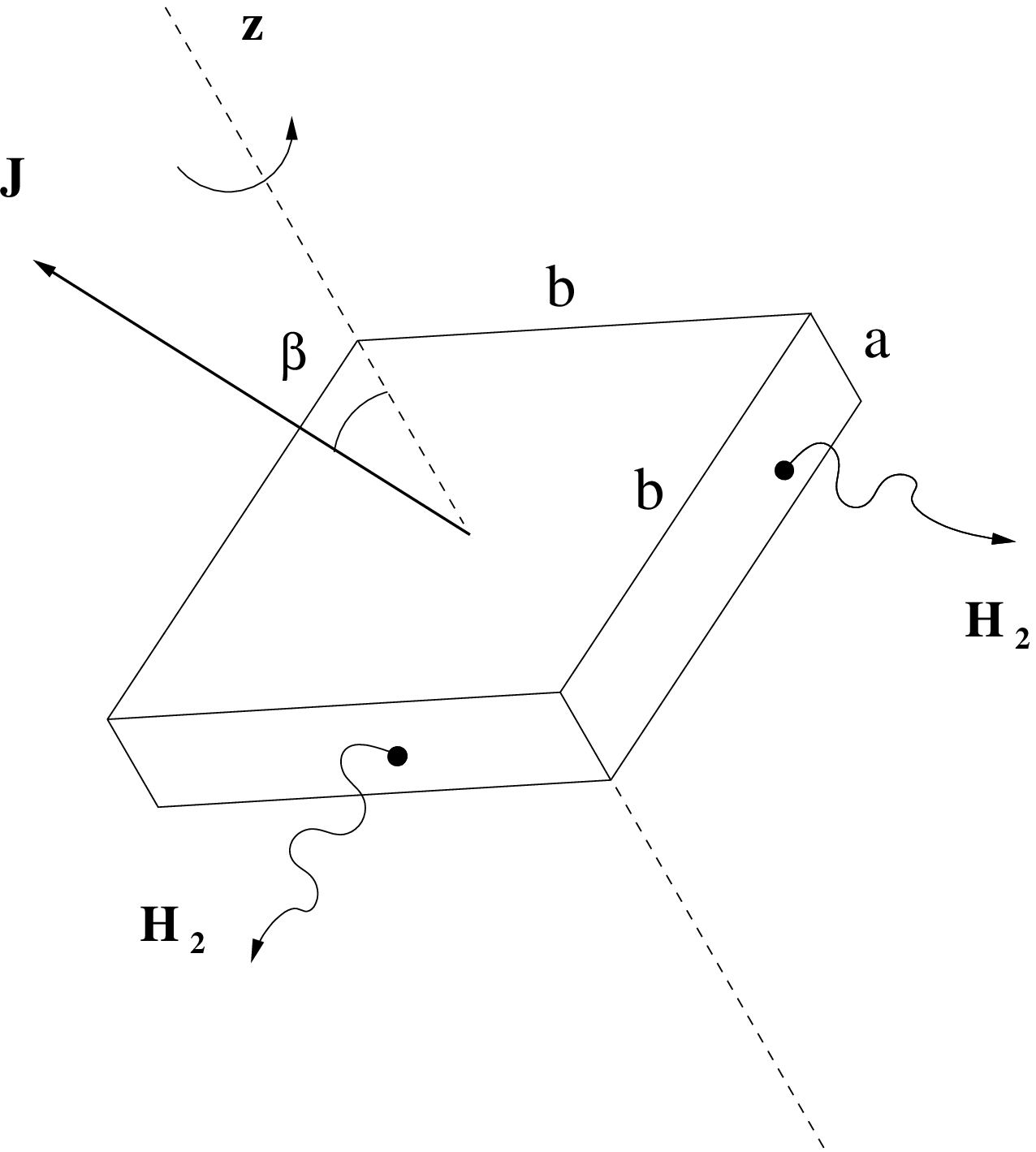} 
\caption{{\it Left panel}-- Grain alignment implies several alignment
processes acting simultaneously and spanning many time scales
(shown for $10^{-5}$~cm grain in cold interstellar gas). The
rotational dynamics of a grain is rather complex. The internal alignment
introduced by Purcell (1979) was thought to be slower than precession
until Lazarian \& Draine (1999b, henceforth LD99b) showed that it happens $10^6$ times
faster when relaxation through induced by nuclear spins is accounted for
(approximately $10^4$~s for the $10^{-5}$~cm grains). {\it Right panel}--
Grain rotation arising from systematic torques arising from H$_2$ formation
(P79). In the presence of efficient internal relaxation the angle $\beta$
between the axis of maximal moment of inertia and $\bf J$ is small is
grain is rotating at suprathermal rates ($E_{kinetic}\gg kT_{grain}$).}
\end{figure}

A very different process of grain spin-up can be found in a very important, 
but not timely appreciated
work by Dolginov \& Mytrophanov (1976).  These authors considered 
differential scattering of
photons of right and left circular polarization by an irregular dust grain. 
As the size of the irregularities
gets comparable with the wavelength, it is natural that interaction of a 
grain with photons will depend
on the photon polarization. Unpolarized light can be presented as a 
superposition of equal number
of left and right circularly polarized photons. Therefore it is clear 
that the interaction with photons
of a particular polarization would deposit angular momentum to the grain. 
The authors concluded that for typical diffuse ISM conditions this process 
should induce grain rotation at 
suprathermal velocities. However, while Purcell's torques became a textbook 
stuff, radiative 
torques had to wait 20 years before they were 
reintroduced to the field (Draine 1996, Draine \& Weingartner 1996, 1997).  

It was realized by Martin (1971) that rotating charged grains will develop
magnetic moment and the interaction of this moment with the interstellar
magnetic field will result in grain precession.  
However, soon  a process that
renders much larger magnetic moment was discovered (Dolginov \& Mytrophanov 
1976). This process is the 
Barnett effect, which is converse of the Einstein-de Haas effect.
If in Einstein-de Haas effect a paramagnetic body starts rotating
 during remagnetizations
as its flipping 
electrons transfer the angular momentum (associated with their spins)
 to the
lattice, in the Barnett effect
the rotating body shares its angular momentum with the electron
subsystem  causing magnetization. The magnetization
is directed along the grain angular velocity and the value
of the Barnett-induced magnetic moment is $\mu\approx 10^{-19}\omega_{(5)}$~erg
gauss$^{-1}$ (where $\omega_{(5)}\equiv \omega/10^5{\rm s}^{-1}$). Therefore
the Larmor precession has a period 
$t_{Lar}\approx 3\times 10^6 B_{(5)}^{-1}$~s. If magnetic field direction
changes over timescales much larger than $t_{Lar}$, the orientation of
grain angular momentum and magnetic field is preserved. Thus MHD turbulence
in molecular clouds (see Lazarian \& Cho 2004) does not destroy grain alignment. This fast Larmor precession makes magnetic field in most cases 
the axis of alignment.

Being solid bodies, interstellar grains can rotate about 3 
different principal axes of grain inertia. As the
result they tumble while rotating. This effect was attracting attention 
of the early researchers (see
Jones \& Spitzer 1967) till Purcell (1979) identified internal relaxation 
within grains as the process
that can suppress grain rotation about all axes, but the axis corresponding 
to the grain maximal
moment of inertial (henceforth axis of maximal inertia). Indeed,  consider 
a spheroidal grain, which
kinetic energy can be presented as (see Lazarian \& Roberge 1997) 
\begin{equation}
E(\theta)=\frac{J^2}{I_{max}}\left(1+\sin^2\beta (h-1)\right), 
\end{equation}
where $\beta$ is the angle between the axis of major inertia and grain 
angular momentum (see Fig.~1).
In the absence of external torques grain angular momentum is preserved. 
The minimum of grain energy 
corresponds therefore to $\beta=0$, or grain rotating exactly about the 
axis of maximal inertia. As 
internal dissipation decreases kinetic energy, it sounds natural that
 $\beta=0$ is the expected 
state of grain subjected to fast internal dissipation. 

\section{What is the physics of internal alignment?}

Purcell (1979) introduced a new process of internal dissipation which he 
termed "Barnett relaxation". 
This process may be easily understood. We know that a 
freely rotating grain preserves the direction of
${\bf J}$, while angular velocity precesses about 
${\bf J}$.
We learned earlier that the Barnett effect results in the magnetization
vector parallel to $\vec \Omega$. As a result, the Barnett magnetization
will precess in body axes and cause paramagnetic dissipation.
The ``Barnett equivalent magnetic field'', i.e. the equivalent external
magnetic field that would cause the same magnetization of the grain  
material, is $H_{BE}=5.6 \times10^{-3} \omega_{(5)}$~G, 
which is much larger than the interstellar magnetic 
field. Therefore the Barnett relaxation happens on the scale $t_{Bar}\approx
4\times 10^7 \omega_{(5)}^{-2}$~sec,
i.e. essentially instantly compared to the time that it takes to
damp grain rotation for typical molecular cloud conditions.

Even stronger relaxation process has been identified recently by 
Lazarian \& Draine (1999a). They termed it ``nuclear relaxation''.
 This is an analog of Barnett
relaxation effect that deals with nuclei. Similarly to unpaired electrons
nuclei tend to get oriented in a rotating body. However the nuclear analog
of ``Barnett equivalent'' magnetic field is much larger and Lazarian \&
Draine (1999a) concluded that the nuclear relaxation can be a million times
faster than the Barnett relaxation. 

Why would the actual relaxation rate matter? The
rate of internal relaxation couples grain rotational and vibrational
degrees of freedom. LD99b showed that this
will result in grain ``thermal flipping''. Such a flipping would
average out Purcell's torques and result in grain being
``thermally trapped'' in spite of the presence of uncompensated
torques. Whether a grain gets ``thermally trapped'' depends on
its size (with the grains less than a critical size $a_c$
rotating thermally). 
While Barnett and inelastic relaxation (see also Lazarian \& Efroimsky
1999) results in $a_c$ equal or less than $10^{-5}$~cm, the
nuclear internal relaxation provides $a_c\sim 10^{-4}$~cm. This means
that most grains rotate thermally in the presence of Purcell's torques.
The exception to this thermallization are radiative torques that are
not fixed in grain coordinates. Such torques can spin-up dust in spite
of thermal flipping.

\section{What does align angular momentum of grains?}

While a number of processes can result in grain angular momentum alignment  (see Lazarian 2003), we shall briefly discuss only 3 of them. 

{\bf Paramagnetic Alignment.} ---
Davis-Greenstein (1951)
mechanism (henceforth D-G mechanism)
is based on the paramagnetic dissipation that is experienced
by a rotating grain. Paramagnetic materials contain unpaired
electrons which get oriented by the interstellar magnetic field ${\bf B}$. 
The orientation of spins causes
grain magnetization and the latter 
varies as the vector of magnetization rotates
 in grain body coordinates. This causes paramagnetic loses 
at the expense of grain rotation energy.
Note, that if the grain rotational velocity ${\vec \Omega}$
is parallel to ${\bf B}$, the grain magnetization does not change with time
and therefore
no dissipation takes place. Thus the
paramagnetic dissipation  acts to decrease the component of ${\vec \Omega}$
perpendicular to ${\bf B}$ and one may expect that eventually
grains will tend to rotate with ${\vec \Omega}\| {\bf B}$
provided that the time of relaxation $t_{D-G}$ is much shorter than  $t_{gas}$,
the
time of randomization through chaotic gaseous bombardment.
In practice, the last condition is difficult to satisfy. For $10^{-5}$ cm 
grains
in the diffuse interstellar medium
$t_{D-G}$ is of the order of $7\times 10^{13}a_{(-5)}^2 B^{-2}_{(5)}$s , 
while  $t_{gas}$ is $3\times 10^{12}n_{(20)}T^{-1/2}_{(2)} a_{(-5)}$ s (
see table~2 in Lazarian \& Draine 1997) if
magnetic field is $5\times 10^{-6}$ G and
temperature and density of gas are $100$ K and $20$ cm$^{-3}$, respectively. 
However, at the time when it was introduced ,in view of uncertainties in
interstellar parameters, the D-G mechanism looked plausible.

The first detailed analytical treatment of the problem of D-G
alignment was given by Jones \& Spitzer (1967) who described the alignment
of ${\bf J}$
using a Fokker-Planck equation. This 
approach allowed them to account for magnetization fluctuations
within grain material and thus provided a more accurate picture of 
${\bf J}$ alignment. 
The first numerical treatment of
D-G alignment was presented by Purcell (1969). 
By that time it became clear that the D-G
mechanism is too weak to explain the observed grain alignment. However,
Jones \& Spitzer (1967) noticed that if interstellar grains
contain superparamagnetic, ferro- or ferrimagnetic (henceforth SFM) 
inclusions\footnote{The evidence for such inclusions was found much later
through the study of interstellar dust particles captured in
the atmosphere (Bradley 1994).}, the
$t_{D-G}$ may be reduced by orders of magnitude. Since $10\%$ of
atoms in interstellar dust are iron
the formation of magnetic clusters in grains was not far fetched
(see Martin 1995).
However, detailed calculations in Lazarian (1997), Roberge \& Lazarian
(1999) showed that the alignment achievable cannot account for
observed polarization coming from molecular clouds provided
that dust grains rotate thermally. This is the consequence of
thermal fluctuations within grain material. These internal
magnetic fluctuations
randomize grains orientation in respect to magnetic field if
grain body temperature is close to the rotational temperature.

Purcell (1979) pointed out that fast rotating grains are immune to
both gaseous and internal magnetic 
randomization. Thermal trapping limits the range of grain sizes
for which Purcell's torques can be efficient (Lazarian \& Draine 1999ab).
For grains that are less than the critical size, which can be $10^{-4}$~cm
and larger, rotation is essentially thermal. Alignment of such grains
is expected in accordance with the DG mechanism predictions (see 
Roberge \& Lazarian 1999) and seem to be able to explain the residual alignment
of small grains that is seen in the Kim \& Martin (1995) inversion.
An important feature of this weak alignment is that it is proportional to
the energy density of magnetic field. This potentially opens a way for
a new type of magnetic field diagnostics. 

Lazarian \& Draine (2000) predicted
that PAH-type particles can be aligned paramagnetically due to the relaxation
that is faster than the DG process. In fact, they showed that the DG
alignment is not applicable to very fast rotating particles, for which
Barnett magnetic field gets comparable with magnetic fields of the atom
neighbors.

{\bf Mechanical Alignment.} ---
Gold (1951) mechanism is a process of mechanical alignment of grains. Consider
a needle-like grain interacting with a stream of atoms. Assuming
that collisions are inelastic, it is easy to see that every
bombarding atom deposits angular momentum $\delta {\bf J}=
m_{atom} {\bf r}\times {\bf v}_{atom}$ with the grain, 
which is directed perpendicular to both the
needle axis ${\bf r}$ and the 
 velocity of atoms ${\bf v}_{atom}$. It is obvious
that the resulting
grain angular momenta will be in the plane perpendicular to the direction of
the stream. It is also easy to see that this type of alignment will
be efficient only if the flow is supersonic\footnote{Otherwise grains
will see atoms coming not from one direction, but from a wide cone of
directions (see Lazarian 1997a) and the efficiency of 
alignment will decrease.}.
Thus the main issue with the Gold mechanism is to provide supersonic
drift of gas and grains. Gold originally proposed collisions between
clouds as the means of enabling this drift, but later papers (Davis 1955) 
showed that the process could  only align grains over limited patches of
interstellar space, and thus the process
cannot account for the ubiquitous grain 
alignment in diffuse medium.

Suprathermal rotation introduced in Purcell (1979) persuaded researchers
that mechanical alignment is marginal. Indeed, fast rotation makes
it difficult for gaseous bombardment to align grains. However, two new
developments must be kept in mind. First of all, it has been proved that mechanical alignment of suprathermally rotating grains
is possible (Lazarian 1995, Lazarian \& Efroimsky 1996, Efroimsky 2002). Moreover, recent work on
grain dynamics (Lazarian \& Yan 2002, Yan \& Lazarian 2003) has shown
that MHD turbulence can render grains with supersonic velocities.
While we do not believe that mechanical alignment is the dominant
process, it should be kept in mind while analyzing observations
(see Rao et al. 1998).

{\bf Alignment via Radiative Torques.} ---
Anisotropic starlight radiation can both spin the grains and align them.
This was first realized by Dolginov \& Mytrophanov (1976), but this
work definitely 
came before its time. The researchers did not have reliable means
to study dynamics of grains and the impact of their work was marginal.
Before Bruce Draine realized that the torques
can be treated with the versatile discrete dipole approximation (DDA)
code ( Draine \& Flatau 1994) the radiative torque alignment was
very speculative. For instance, earlier on
difficulties associated with the analytical approach to
the problem were discussed in Lazarian (1995).
However, very soon after that Draine (1996) modified the DDA code
 to calculate the torques acting on grains of arbitrary
shape. His work revolutionized the field! 
The magnitude of torques were found to be substantial and present
for grains of various irregular shape (Draine 1996, Draine \& Weingartner
1996). After that it became impossible
to ignore radiative torque alignment.

One of the problem of the earlier treatment was that in the presence
of anisotropic radiation the torques will change as the grain aligns
and this may result in a spin-down.  Moreover,
anisotropic flux of radiation will deposit angular momentum 
which is likely to overwhelm rather weak paramagnetic torques. These sort of
questions were addressed by Draine \& Weingartner (1997) and it was
found that for most of the tried grain shapes the torques tend to 
align ${\bf J}$ along magnetic field. The reason for that is yet unclear
and more work is clearly necessary before we can treat 
radiative alignment as a theory rather than an empirical fact. One
of the authors of the review (AL) recalls that this was also the opinion
of Lyman Spitzer who got interested in the action of radiative torques
and was encouraging the author to do analytical work and simple testing
to clarify the essence of the radiative torque alignment. 
One of the missing pieces of physics, namely the dynamics of radiative
torques,  has been dealt with recently by Weingartner \&
Draine (2003), who treated flipping of grains in the presence of monochromatic
radiation. 

{\bf Quantitative description.} ---
As we discussed above, to relate the polarization to magnetic field, the
Rayleigh reduction factor $R$ should be calculated. This factor was calculated
for DG alignment (see Roberge \& Lazarian 1999), for Purcell alignment (see Lazarian \& Draine 1997), mechanical alignment of thermally (see Lazarian 1997a)
and suprathermally rotating grains (see Lazarian 1995, Lazarian \& Efroimsky 1996,
Efroimsky 2002). For radiative torques no quantitative theory exists. An
educated guess may be that for grains larger than the critical size $R=1$,
i.e. the grains are perfectly aligned. The calculations of the critical
grain size can be done by comparing the radiative torques calculated with
the DDA software and the damping of grain rotation via gas-grain, ion-grain
collisions, plasma drag etc. (see Draine \& Lazarian 1999 for a description
of various damping mechanisms).

\section{ What can align grains deep in molecular clouds?}

We believe  that a substantial degree
of understanding of grain alignment has been achieved recently. 
For the first time ever the available observational data
look consistent with the theoretical expectations. 

Both the dependences of the polarization degree versus wavelength
that follow Serkowski law (Serkowski 1973) and 
studies
of changes of polarization degree with the wavelength done in Far Infrared
(see Hildebrand
2000) are consistent with theoretical predictions (see discussion in
Lazarian, Goodman \& Myers 1997, henceforth LGM97). 
According to Lazarian (2003) the study of grain alignment
at the diffuse/dense cloud interface by Whittet et al. (2001)
is suggestive that grains are aligned by radiative torques. 
Indeed, the latter study finds that the properties of grains stay
the same, while the minimal size of the aligned grains is
increasing with the increase of extinction. This behavior is
inconsistent with superparamagnetic
grains discussed in Mathis (1986). For those grains the size of aligned
grains is determined by the presence of superparamagnetic inclusion and
does not change unless the grain size distribution changes. On the contrary,
radiative torque efficiency decreases for smaller grains as the shorter wavelength radiation
field gets preferentially attenuated by extinction.

\begin{figure*}[h!t]
\includegraphics[width=.48\textwidth]{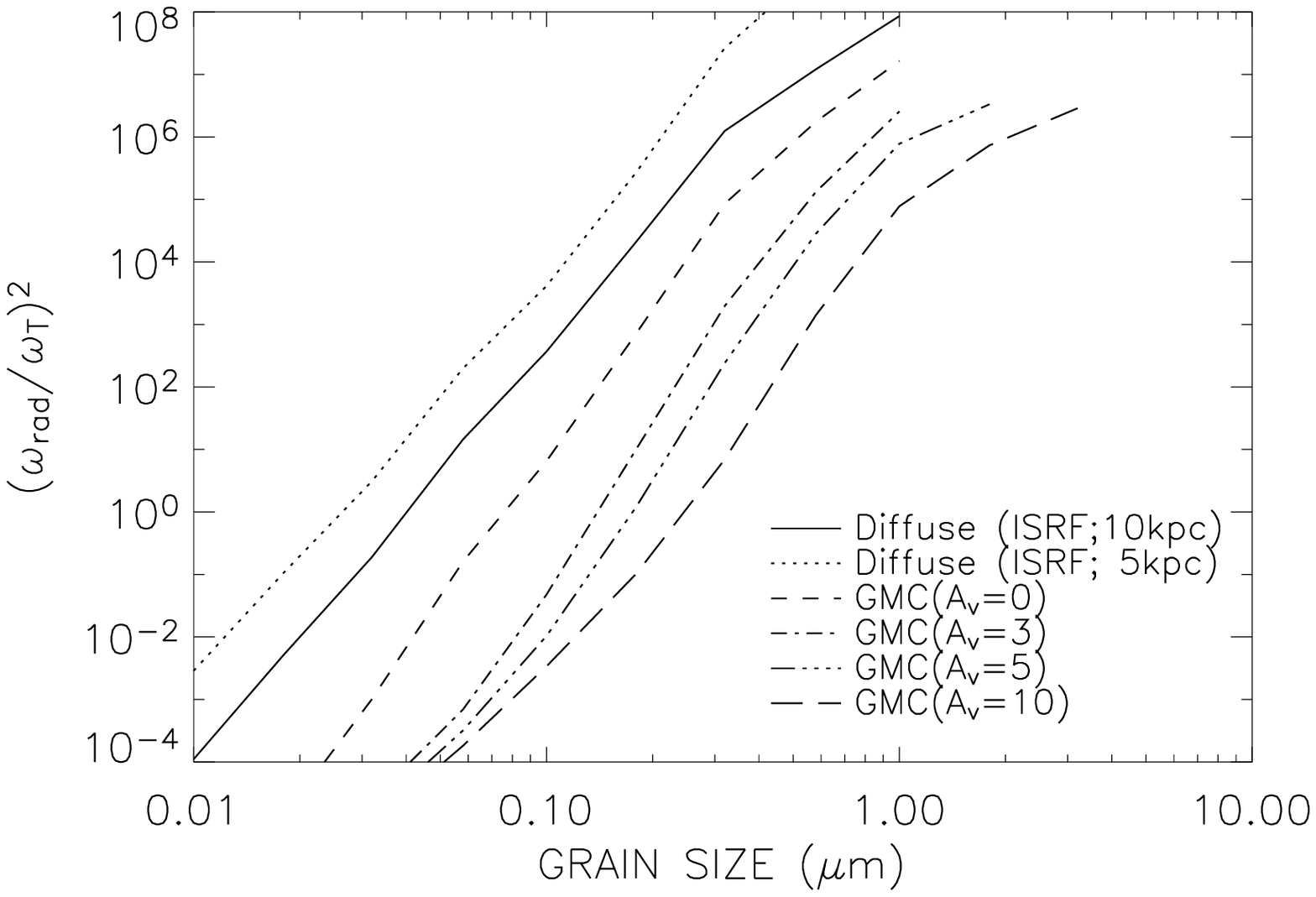}
\includegraphics[width=.46\textwidth]{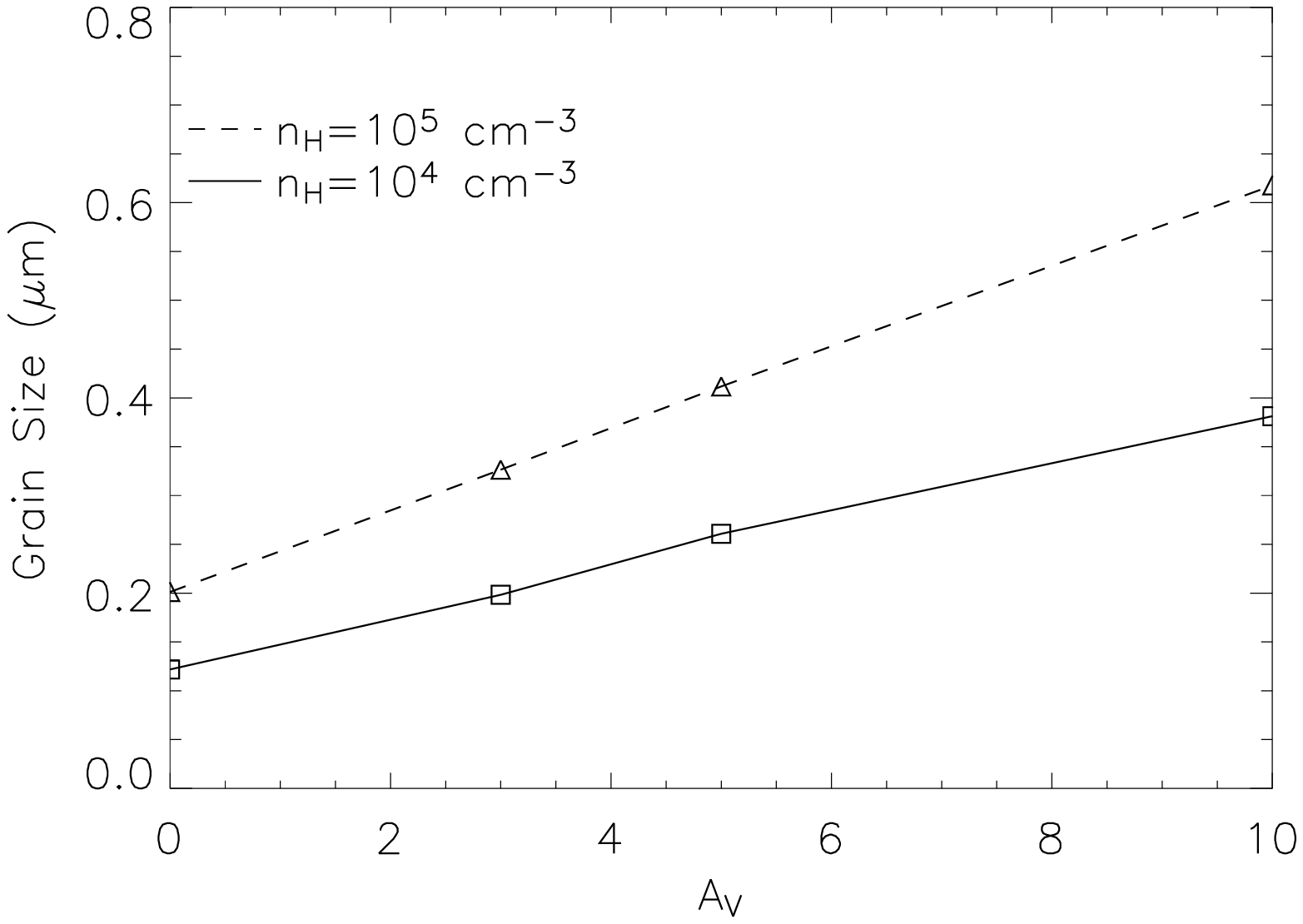}
\caption{
    Radiative torque at high extinction by Cho \& Lazarian (2004). 
(a) Different curves represent radiative
    torque by an anisotropic part of 
    radiation field. The degree of anisotropy of
10 \% was assumed for ISRF. The visual extinction $A_V$
    is for a giant molecular cloud located at 5kpc from
    the Galactic center. Although 
    the UV smoothed refractive index of silicon is used for
small grains, our results are consistent with torques published
in Draine \& Weingartner (1996).  
    (b) Aligned grain size vs. visual extinction $A_V$. For the threshold
suprathermal angular velocity 5 times larger than the thermal angular
velocity was chosen. It is clear that increase of grain size can compensate
for the extinction of light in cloud cores.
 Solid line: $n_H=10^4 cm^{-3}$; Dotted line: $n_H=10^5 cm^{-3}$ in the cloud.
}
\end{figure*}

An earlier review of observational molecular cloud data was given in
LGM97. 
It broadly reconciled the near-infrared
data that was suggestive of the suppression of grain alignment at high
extinction and the far-infrared data suggestive of grain alignment
in the vicinity of stars deep embedded into molecular cloud.
LGM97 showed that within molecular clouds far from embedded stars
all the grain alignment mechanisms fail, while near the stars a few
of them, particular radiative torques looked promising.

Data summarized in Hildebrand (2003) suggest that either
hot grains in the vicinity of stars or cold grains at the cloud boundary
are well aligned, while the warm grain at the bulk of the cloud
are marginally aligned. This data are consistent with the LGM97 expectations.
However, the data obtained for pre-stellar cores in Ward-Thompson et al. 
(2000) at the first glance seem to be at odds with the LGM97 predictions. 
Indeed, the properties of these cores summarized in Ward-Thompson et al.
(2002) and Crutcher et al. (2004) fit into the category of zones that
must be dead for grain alignment according to LGM97. 

What could be wrong with LGM97 arguments? The latter paper treats grains
of $10^{-5}$~cm size. Such grains are typical for diffuse ISM, while
grains in prestellar cores can be substantially larger. Grain alignment is
a function of size. Therefore the estimates in LGM97 should be reevaluated.

Cho \& Lazarian (2004, preprint, henceforth CL04) revealed a steep dependence of 
radiative torque 
efficiency on grain size. While an earlier study by Draine \& Weingartner
(1996) was limited by grains with size $a \leq 0.2 \times 10^{-4}$~cm, 
CL04 studied grains up
to $3\times 10^{-4}$~cm size subjected to the attenuated radiative field calculated
in accordance with the prescriptions in Mathis, Mezger \& Panagia (1983).
Figure~3 shows that large grains can be efficiently span up by radiative
torques even at the extinction of $A_v$ of 10 and higher. Real molecular
clouds are likely to be inhomogeneous. As the result, the radiation has
more chances to penetrate deep within molecular clouds\footnote{
Even larger grains are known to be present in the accretion discs around
young stars. Grain alignment may be efficient for such grains revealing the
structure of the all-important magnetic fields. However, this issue
is beyond the scope of this review.}.

In general, alignment of large  grains by other mechanisms can also be more
efficient. Such grains are not subjected to thermal trapping (see Lazarian \& Draine 1999ab) and therefore can be aligned by Purcell's mechanism (see
Lazarian \& Draine 1997 calculations that take into account crossover dynamics).
Larger grains also get larger velocities due to turbulent motions (see
Yan \& Lazarian 2003) and therefore are more likely to be aligned mechanically.
This gives further hope that using Far Infrared polarimetry it is possible 
to trace magnetic fields deep in molecular clouds.

\section{Summary}

~~~~~1. Aligned grains provide a unique way to study magnetic field. As we better understand grain alignment the interpretation of
emission and absorption polarization data in terms of underlying magnetic
field gets more reliable.

2. Grain alignment is a complex process that includes precession and gradual
alignment of angular momentum
in respect to magnetic field and the alignment of grain axes in respect to angular momentum. The latter alignment influences the former one. Rapid precession of
grain angular momentum about magnetic field makes magnetic field the axis alignment even if the alignment mechanism is not of magnetic nature.

3. Internal relaxation is the process that minimizes grain kinetic energy for a fixed
angular momentum. As the result of the process grain rotates about its axis of maximal
inertia. Relaxation related to the nuclear moments within a grain have 
been recently identified as the major mechanism of internal relaxation. 

4. Internal relaxation couples rotational and vibrational degrees of freedom.
 As the result, thermal fluctuations
in grain material prevent perfect alignment of grain axes in respect to angular momentum. Moreover, thermal fluctuations cause rapid flipping and "thermal trapping"
of 
sufficiently small grains, ine. they prevent the grains from
spinning rapidly (suprathermally) even in the presence of uncompensated Purcell's torques. 

5. Paramagnetic alignment is definitely present for small thermally trapped 
grains. However, quantitative theories predict that the degree of expected
alignment is rather marginal and depends on the magnetic field intensity.
Purcell's paramagnetic alignment of suprathermally rotating grains is applicable
only to sufficiently large $>10^{-4}$~cm grains, i.e. to the grains that are
not thermally trapped. The mechanical alignment should not be disregarded
as grains can be driven by turbulence to supersonic velocities. 

6. Radiative torques mechanism is the most promising mechanism for alignment of
grain angular momentum. The efficiency of radiative torques depends on grain size
and properties of ambient radiation field.
This allows to explain why some interstellar grains are aligned,
while others are not aligned. 

7. Alignment of large dust grains is possible within cores of molecular clouds.
Radiative torques efficiency increases substantially for larger grains. As the
result, even substantially attenuated interstellar radiation field can provide
good alignment. 
This makes Far Infrared Polarimetry an essential tool for getting insight into
the magnetic fields in hotbeds of star formation.
          
{\bf Acknowledgments.} This work is supported by the NSF grant AST-0243156.
Help by Ms. H. Yan is acknowledged.

%
%
%

%


\end{document}